\documentclass[%
 aip,
 amsmath,amssymb,
 reprint,%
]{revtex4-1}

\usepackage{graphicx}
\usepackage{dcolumn}
\usepackage{bm}

\usepackage[utf8]{inputenc}
\usepackage[T1]{fontenc}
\usepackage{mathptmx}
\usepackage{etoolbox}

\makeatletter
\def\@email#1#2{%
 \endgroup
 \patchcmd{\titleblock@produce}
  {\frontmatter@RRAPformat}
  {\frontmatter@RRAPformat{\produce@RRAP{*#1\href{mailto:#2}{#2}}}\frontmatter@RRAPformat}
  {}{}
}%
\makeatother
\begin{document}

\preprint{AIP/123-QED}

\title{Achromatic, planar Fresnel-Reflector for a Single-beam Magneto-optical Trap}
\author{S. A. Bondza}
\email[]{saskia.bondza@ptb.de}
\affiliation{Physikalisch-Technische Bundesanstalt, Bundesallee 100, Braunschweig, Germany}
\affiliation{Deutsches Zentrum für Luft- und Raumfahrt e.V. (DLR), Institut für Satellitengeodäsie und Inertialsensorik, Callinstraße 30b, 30167 Hannover, Germany}

\author{T. Leopold}
\affiliation{Physikalisch-Technische Bundesanstalt, Bundesallee 100, Braunschweig, Germany}
\affiliation{Deutsches Zentrum für Luft- und Raumfahrt e.V. (DLR), Institut für Satellitengeodäsie und Inertialsensorik, Callinstraße 30b, 30167 Hannover, Germany}

\author{R. Schwarz}
\affiliation{Deutsches Zentrum für Luft- und Raumfahrt e.V. (DLR), Institut für Satellitengeodäsie und Inertialsensorik, Callinstraße 30b, 30167 Hannover, Germany}

\author{C. Lisdat}
\affiliation{Physikalisch-Technische Bundesanstalt, Bundesallee 100, Braunschweig, Germany}

\date{\today}

\begin{abstract}
We present
a novel
achromatic, planar, periodic mirror structure for single-beam magneto-optical trapping and
demonstrate its use in first- and second-stage cooling and trapping for different isotopes of strontium.
We refer to it as Fresnel MOT as the structure is inspired by Fresnel lenses. By design, it avoids many of the problems that arise for multi-color cooling using planar structures based on diffraction gratings,
which have been the dominant planar structures to be used for single-beam trapping thus far.
In addition to a complex design process and cost-intensive fabrication, diffraction gratings suffer from their inherent chromaticity,
which causes different axial displacements of trap volumes for the different wavelengths and necessitates tradeoffs in their diffraction properties and achievable trap depths.
In contrast, the Fresnel reflector structure presented here is a versatile, easy-to-manufacture device that combines achromatic beam steering with the advantages of a planar architecture.
It enables miniaturizing trapping systems for alkaline-earth-like atoms with multiple cooling transitions as well as multi-species trapping in the ideal tetrahedral configuration and within the same volume above the structure.
Our design presents a novel approach for the miniaturization of cold-atom systems based on single-beam MOTs and enables the widespread adoption of these systems.
\end{abstract}

\maketitle


\maketitle

\section{Introduction}
The invention of laser cooling and trapping techniques
has enabled control and manipulation of individual quantum systems such as ensembles of atoms and heralded the emergence of quantum sensing based on cold atoms \cite{deg17}.
This field of research continues to improve measurement resolution 
and to revolutionize measurement capabilities in many fields, including metrology \cite{eck18, mcg19, rie15, sch18c}, navigation and inertial sensing \cite{alz19} as well as the search for new physics \cite{saf18a}.
Optical lattice clocks \cite{lud15} in frequency metrology, matter-wave interferometers \cite{pet99, kas92} for gravimetry, and (cold-atom-based) magnetometers\cite{ven07, ock13, wil05a} are only some of the new measurement devices that are based on cold atoms. Additionally, cold atoms are used in quantum computing experiments \cite{kau15, pic18, saf16, hen20}.
Due to their high accuracy, precision, and vast range of applications,
cold-atom quantum sensors are transitioning from laboratory environments to field-based applications, including air- and spaceborne platforms \cite{tin07}. The resulting need for (trans)portability places stringent requirements on a sensor's size, weight, and power consumption (SWaP) \cite{bon19}. Miniaturizing the key components of quantum experiments to harness the capabilities of state-of-the-art quantum sensors in field operation is thus a vital, ongoing process \cite{mcg22}.
For neutral atoms, preparation of an ensemble for a measurement usually starts with a magneto-optical trap (MOT).
A number of more compact alternatives to the traditional six-beam geometry\cite{raa87} have been developed throughout the years, including the ``pyramid MOT'' \cite{lee96a, bow19a}, ``tetrahedral MOT'' \cite{van09}, and ``grating MOT'' (gMOT) \cite{mcg16, mcg17, nsh13}.\par
Grating MOTs are based on a planar, monolithic diffraction grating and thus offer the highest degree of miniaturization as well as unrestricted radial access.
They have been employed in several quantum sensors using alkaline atoms already \cite{elv19, bar19b, lee21}.
Here, they are well suited as alkaline atoms are usually cooled using a single transition with a linewidth of several megahertz and highly effective sub-Doppler cooling. 
However, many applications require multi-color operation with wavelengths separated by up to several hundred nanometers, including the multi-stage cooling of alkaline-earth-like atoms, which is used to reach temperatures of few microkelvin \cite{kat99, kuw99, doe13}, as well as multi-species cooling and trapping.
Alkaline-earth-like atoms are particularly useful for many metrological and sensing applications such as optical clocks \cite{lud15, hu17a} due to the presence of ultranarrow optical transitions.
For these applications, the inherent chromacity of diffraction gratings poses a number of challenges to cooling and trapping at two vastly different wavelengths: Firstly, the respective trapping volumes are displaced axially, which complicates the transfer between MOT stages using different wavelengths \cite{bon22}.
Secondly, the grating needs to be specifically designed and manufactured for each atomic species or combination of species \cite{bur23}, which is a costly, time-consuming, and complex process and requires micro-fabrication techniques. 
Lastly, achieving efficient diffraction at both wavelengths may require limiting the diffraction angle $\theta$, e.g., to $\theta<40$° for the 461~nm transition in Sr.
This results in significantly decreased trap depths as large diffraction angles benefit the trap depth \cite{van11b, bar23}.
Additional compromises are often necessary with regards to diffraction efficiencies and polarization purity, which further impacts trap dynamics.\par
Two-stage cooling in a grating MOT has been demonstrated for the bosonic isotope $^{88}$Sr \cite{bon22, elg22a}.
However, fermionic alkaline-earth atoms are often more challenging to cool and exhibit an increased sensitivity to the trap geometry \cite{bar23}, because the hyperfine splitting of their energy levels and their substantially different Land\'{e} factors in the ground and excited state result in more complex cooling dynamics \cite{muk03}.\par
\begin{figure}
	\includegraphics[width=\linewidth]{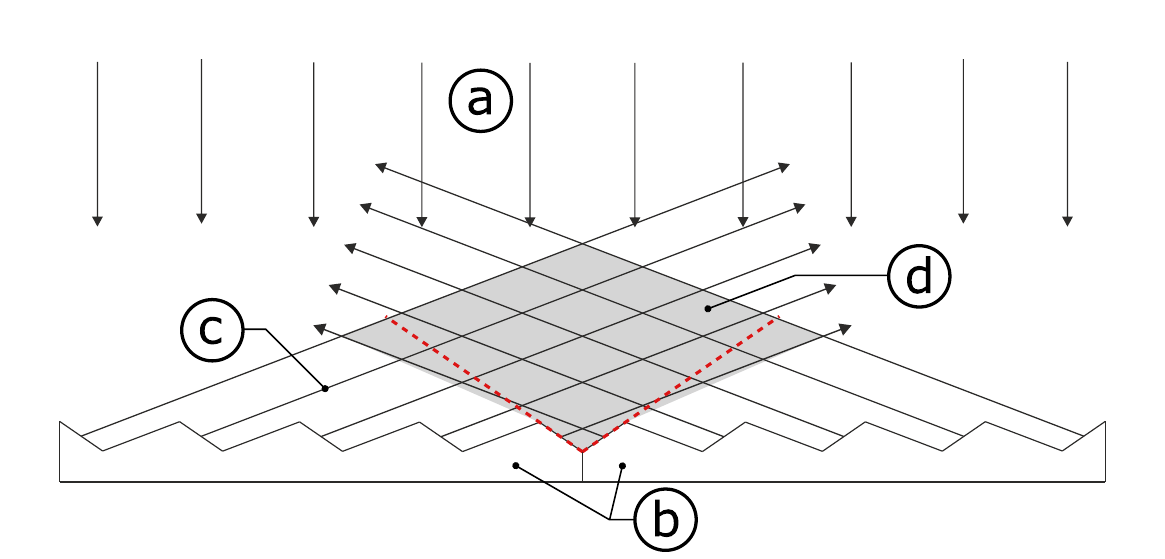}%
	\caption{Schematic of the beam propagation in the Fresnel MOT. The incident beam (a) is reflected by the Fresnel reflector (b), creating secondary beams (c). Adding a magnetic quadrupole field with its minimum within the beam overlap volume (d) leads to cooling and trapping of atoms. The red dashed lines show the equivalent mirror surfaces of a tetrahedral MOT producing an identical beam geometry. Figure adapted with permission from S. Bondza \cite{bon23a}. \label{fig:geometry}}
\end{figure}

The pyramid and tetrahedral MOTs, being reflective structures, are inherently achromatic and thus do not suffer from the same problems as grating MOTs for multi-color operation.
They do not offer the same level of optical access and miniaturization, however:
The pyramid MOT conserves the standard six-beam geometry using a number of mirror pairs that reflect the incident beam at a 90° angle \cite{lee96a, bow19a}.
As the trap volume lies within the pyramid, optical access is severely limited.
Furthermore, this set up is relatively bulky compared to the planar chip design of grating MOTs.
In the tetrahedral MOT, an incident beam is reflected by three surfaces arranged such that the incident beam and the reflected beams cross close to the tetrahedral configuration \cite{van09}.
Part of the beams' overlap volume extends above the mirror structure (see Fig. \ref{fig:geometry}, red dashed line).
Still, optical access is severely limited since the mirrors block radial access to more than half the overlap volume.
This can be compensated by pulling the mirrors apart, but only at the cost of reduced laser power density or trapping volume.\par
Here, we present a novel type of single-beam MOT that combines  unrestricted radial access and the compactness of a planar design like the grating MOT with the achromatic beam steering of reflective operation \cite{bon22a}. It is based on the tetrahedral mirror MOT but replaces the bulk mirrors with surface-structured Fresnel-like reflectors. This leads to a quasi-planar, achromatic, single-beam MOT with full optical access in the radial plane providing cooling in tetrahedral geometry. Additionally, the planar geometry enables the integration of further miniaturized components in a compact package, e.g. atom source, magnetic field coils, optical resonators.
It further allows trapping of different atomic species at the same height over the reflector, independent of wavelength.
We demonstrate two-stage trapping and cooling of $^{88}$Sr atoms in the Fresnel MOT. Initially, atoms are cooled on the $^1S_0 \rightarrow {^1P_1}$ transition at a wavelength of $461\,\mathrm{nm}$ and a linewidth of $30.2\,\mathrm{MHz}$ \cite{nic15}. They are then transferred to a so-called broadband MOT on the $^1S_0 \rightarrow {^3 P_1}$ transition at a wavelength of $689\,\mathrm{nm}$ and a linewidth of $7.5\,\mathrm{kHz}$ \cite{yas06} reaching temperatures of $23\,\mu\mathrm{K}-43\,\mu\mathrm{K}$.
We further demonstrate trapping and cooling of $^{87}$Sr and $^{86}$Sr atoms using the $^1S_0 \rightarrow {^1P_1}$ transition.

The surface structure of the Fresnel MOT is obtained when the mirror surfaces of a tetrahedral MOT are folded back onto a common plane as illustrated in Fig.~\ref{fig:geometry}. Surfaces reflecting towards the center of the trap alternate with those reflecting outwards. Ideally, the surfaces of the latter are parallel to the reflected MOT beams ((c) in Fig.~\ref{fig:geometry}) to prevent these beams from reaching the trapping region by multiple reflections. 
The height of the structure is directly related to the period and in principle arbitrary, as long as the period is large compared to the employed wavelength to avoid diffraction. In practice, the period should be chosen as large as conveniently possible, to minimize potential wavefront errors. In the limit of long periods, this eventually leads to the original tetrahedral MOT.\par 
The theory of operation for the Fresnel MOT is identical to the tetrahedral MOT, covered extensively by Vangeleyn et al. \cite{van11b, van09}. A stable optical molasses is produced with a set of beams with a balanced radiation pressure:
\begin{equation}
	\label{eq:balance}
	\sum_{n=1}^{N_{tot}} I_n \vec{k_n} = 0 ,
\end{equation}
where $I_n$ and $\vec{k_n}$ are the intensity and wave vector of beam $n$, respectively and $N_{tot}$ is the total number of beams. In the radial plane, given a radially symmetric formation of secondary beams, this condition is fulfilled when illuminating the structure with a flat-top intensity profile. Axially, the secondary beam intensity can be controlled via the reflectivity $R$ of the Fresnel reflector. For $N$ secondary beams, a balanced radiation pressure is achieved for a mirror inclination with respect to the incident beam of
\begin{equation}
	\label{eq:theta}
	\phi = \frac{1}{2}\arccos\frac{1}{NR}.
\end{equation}
The angle between incident and secondary beams is then given as $\theta=2 \phi$ which corresponds to the convention of the diffraction angle given for grating MOTs.
The minimum number of secondary beams for three-dimensional trapping is $N=3$, which also leads to the largest possible beam overlap volume. In this case, a mirror inclination of up to $\phi = 35.26^{\circ}$, corresponding to the tetrahedral configuration, with a reflectivity of $R=1$ of the mirror surfaces can be chosen.  
The main advantage of the Fresnel MOT over the standard tetrahedral mirror MOT is the possibility for a much more compact, quasi-planar design with radial access as the grating MOT offers. Additionally, strong trapping and radial cooling close to the tetrahedral configuration is maintained independent of wavelength. Fresnel MOTs thus lend themselves for use in miniaturized vacuum systems of a few cubic centimeters volume. 
Apart from light, another critical aspect of MOTs is the generation of sufficiently strong magnetic field which is often generated by a pair of coils in or near anti-Helmholtz configuration. 
In order to minimize their SWaP, placing them as close as possible to the trap volume is mandatory, which is supported by our Fresnel MOT approach. Especially for alkaline-earth atom MOTs, which require a larger magnetic field gradient compared to alkaline atoms, this can lead to designs which forego the common water cooling of the field coils.

\section{Experiment}
\begin{figure}
	\includegraphics[width=0.8\linewidth]{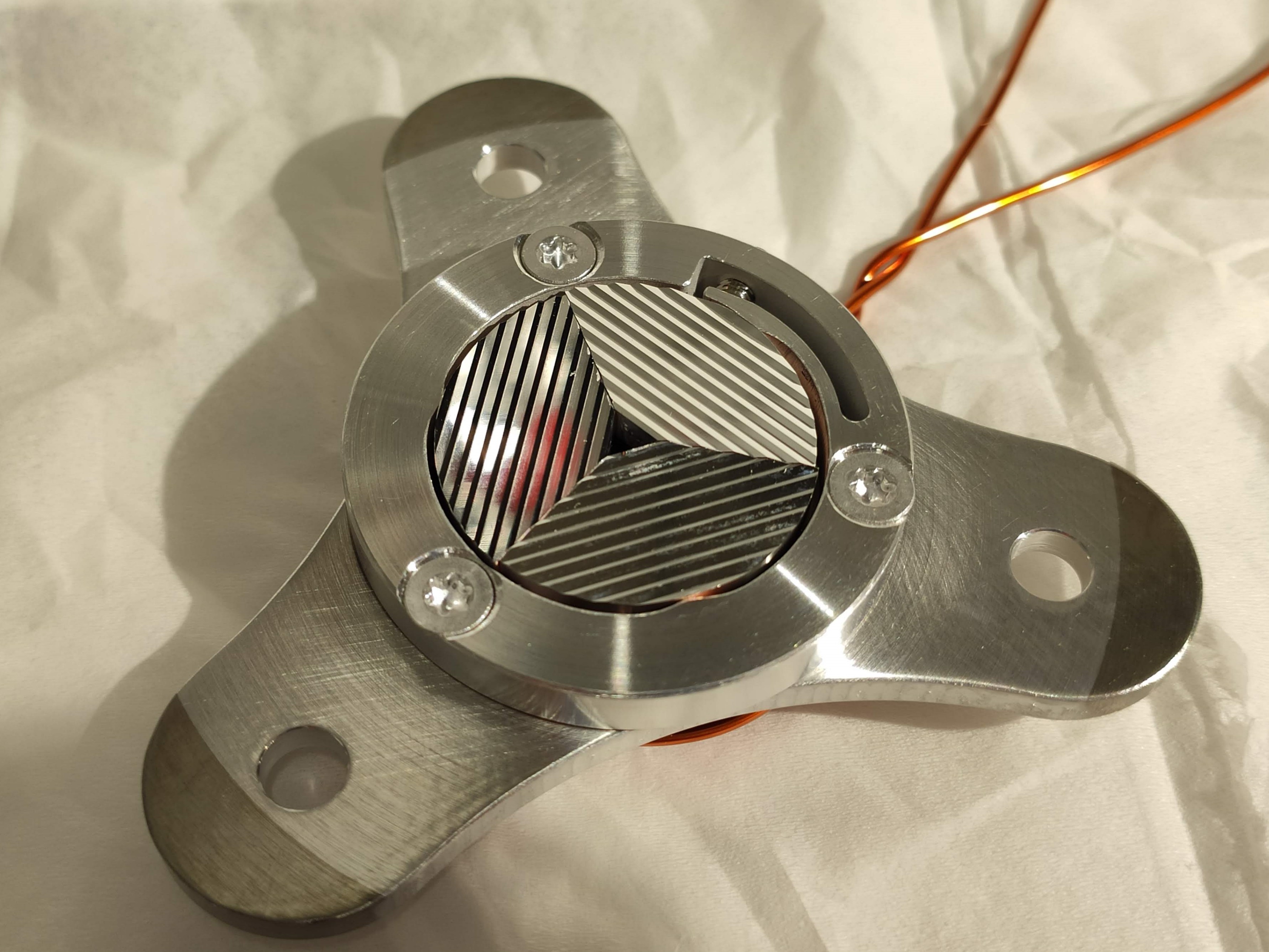}%
	\caption{Photograph of the Fresnel reflector in a 25.4~mm flexure mount.  \label{fig:photo}}
\end{figure}
A prototype of the Fresnel MOT was manufactured from oxygen-free high-conductivity copper. This material is well-suited for this purpose, as a mirror-finish surface can be produced by milling processes alone, without subsequent polishing. Three individual $120\,\mathrm{^{\circ}}$ circle segments where machined that, put together, result in a $25.4\,\mathrm{mm}$ diameter reflector which is depicted in Fig.~\ref{fig:photo}. The copper substrates where sputter-coated with a $100\,\mathrm{nm}$ layer of aluminium that serves as a highly reflective surface in the visible wavelength range. The reflectivity of aluminium in the visible spectrum is about $90\,\%$. This determined the mirror inclination according to equation (\ref{eq:theta}) to 34$^\circ$ yielding an angle between incident and secondary beams of $\theta=68^\circ$. The assembled device features a triangular hole in the center which can potentially be used to load atoms into the trap as shown in Sitaram et al.\cite{sit21}. Polarization conservation of the reflector was measured with a polarimeter and found to be $\approx95\,\%$. The root mean square surface roughness was measured to be $7\,\mathrm{nm}$. \\
The test set-up used to characterize the chip has been presented in detail in a previous publication \cite{bon22}: A bi-chromatic laser beam red-detuned with respect to the $^1$S$_0 \rightarrow ^1$P$_1$ transition of strontium at $461\,\mathrm{nm}$ and the $^1$S$_0 \rightarrow ^3$P$_1$ transition at $689\,\mathrm{nm}$ is circularly polarized and shaped to a flat-top intensity profile with $18\,\mathrm{mm}$ in diameter.
For the first cooling stage, up to $50\,\mathrm{mW}$ of optical power at $461\,\mathrm{nm}$ are incident on the Fresnel reflector, yielding a maximum saturation parameter of $s=0.5$. A set of in-vacuum anti-Helmholtz coils generates a magnetic field gradient of up to $7\,\mathrm{mT/cm}$ along the incident beam axis. The MOT is loaded from an atomic beam parallel to the Fresnel reflector surface, provided by an atomic oven. The oven is typically operated at temperatures ranging between $350\,$°C-$400\,$°C. Some of the atoms are pre-slowed below the MOT capture velocity by a counter-propagating deceleration beam, red-detuned by about $200\,\mathrm{MHz}$. 
For the second cooling stage, up to $15\,\mathrm{mW}$ of power is available corresponding to a saturation parameter of $s\approx1000$. In this stage the magnetic field gradient is reduced to about $0.3\,\mathrm{mT/cm}$. \\
To allow axial positioning control of the trapping center, additional magnetic field coils enable the application of a bias field that offsets the quadropole magnetic field minimum.
As the $^1$S$_0 \rightarrow ^1$P$_1$ transition is not a closed transition, atoms can escape the cooling cycle by decay to either the $^3$P$_0$ or $^3$P$_2$ dark state via the intermediate $^1$D$_2$ state. A  repumping scheme using 679~nm and 707~nm light overcomes this issue \cite{fal11}.\\
To carry out our characterization measurements, laser-induced fluorescence on the $^1$S$_0 \rightarrow ^1$P$_1$ transition is used.
The fluorescence signal is collected by a large numerical aperture lens and imaged either on a CMOS camera with which we carry out atomic temperature measurements or a photo multiplier tube (PMT) to measure MOT lifetimes.

\section{Results}
We typically trap $5 \cdot10^6-8 \cdot10^6$ $^{88}$Sr atoms in the first MOT stage operated on the $461\,\mathrm{nm}$ $^1$S$_0 \rightarrow ^1$P$_1$ transition. We evaluate the MOT regarding lifetime and temperature.
The temperature is evaluated with the time of flight (TOF) method as a function of intensity of the incident MOT beam. Atoms are initially loaded into the Fresnel MOT with MOT beams at $s=0.4$. The intensity is then decreased to the target intensity. Once the atoms are thermalized, the trapping light is switched off and the atomic cloud expands freely.
The diameter of the atom cloud $\sigma$ is observed at time $t$ and the temperature is extracted from $\sigma^2(t) = \sigma_0^2 + (k_B T/ M) \times t^2$. Here, $\sigma_0$ is the initial size of the atomic cloud and $k_B T/ M $ their thermal energy per atomic mass.
Images of the expanding cloud are recorded with fluorescence imaging with an illumination time of $100\,\mu\mathrm{s}$.
\begin{figure}
	\includegraphics[width=\linewidth]{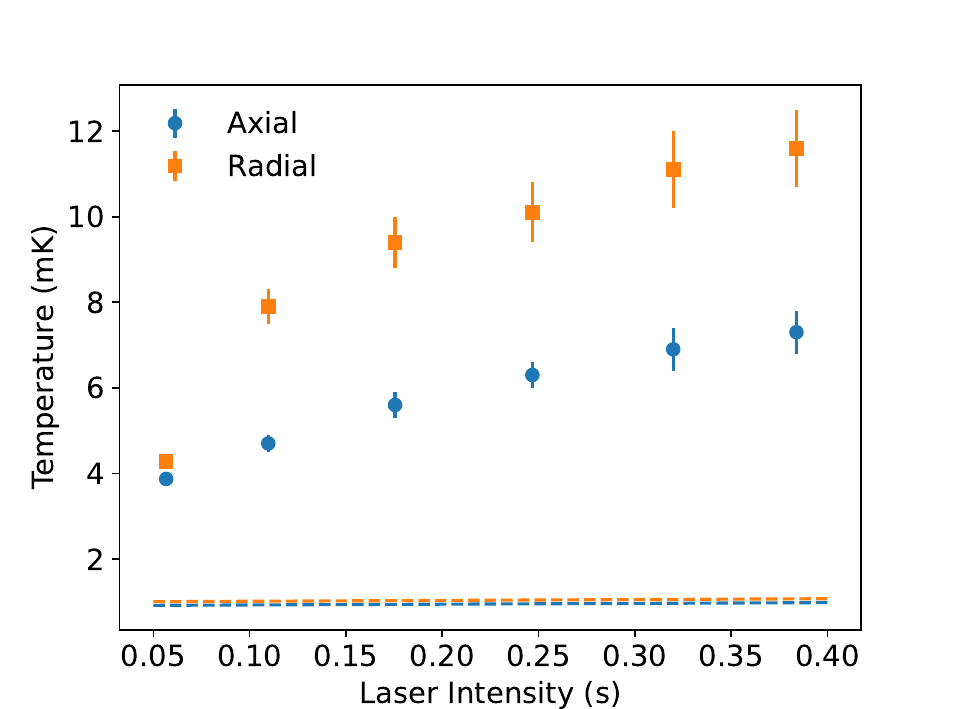}%
	\caption{Axial (blue) and radial (orange) temperature in the first stage MOT as a function of incident laser intensity with $dB/dz = 7\,\mathrm{mT/cm}$. Temperature is measured with the TOF method. Dashed lines indicate the theoretical Doppler limit for both axial and radial direction. Figure adapted with permission from S. Bondza \cite{bon23a}. \label{fig:Temperature_blue}}
\end{figure}

The dependency of temperature on laser intensity is depicted in Fig.~\ref{fig:Temperature_blue}. 
The lowest temperatures are achieved for a saturation parameter of $s=0.06$, with $(3.87\pm0.10)\,\mathrm{mK}$ in the axial direction and $(4.28 \pm 0.12)\,\mathrm{mK}$ in radial direction respectively. For smaller intensities, the atom number becomes too small for temperature measurements due to a high loss rate. These results are comparable with temperatures reached in strontium grating MOTs \cite{sit20, bon22}
and the atoms are sufficiently cold to be transferred to the next cooling stage.\\
The Doppler limit in radial and axial direction as function of the angle $\theta$ is given as $T_{rad} = \frac{T_D}{6} \cdot \left(3 + \frac{1}{\cos(\theta)}\right)$ and $T_{ax} =\frac{T_D}{6} \cdot \left(3 + \frac{1}{\sin^2(\theta/2)}\right)$ respectively \cite{mcg15c}.
The ratio of radial and axial temperature agrees with the expected value of $1.1$  within one standard deviation for the lowest achieved temperatures. 
However, the measured temperatures lie above the theoretical Doppler limit  with an increase much stronger than expected from the Doppler limit. 
This effect is commonly observed in alkaline-earth atoms \cite{xu02, xu03, loo04, oat99} and has been linked to an additional heating mechanism originating from transverse spatial intensity fluctuations\cite{cha05}. At low velocities, cooling is described by a frictional force and an additional constant force resulting from intensity fluctuations. 
Both forces are intensity dependent, the effect thus increases with increasing intensity. 
Possible sources of spatial intensity fluctuations in the Fresnel MOT originate from the beam profile, imperfect edges of neighboring surfaces of the Fresnel structure, imperfections in the coating and slight misalignment of the beam. None of these factors are fundamental limitations and we expect that the performance of the Fresnel MOT can be improved by addressing these issues.\par
\begin{figure}
	\includegraphics[width=\linewidth]{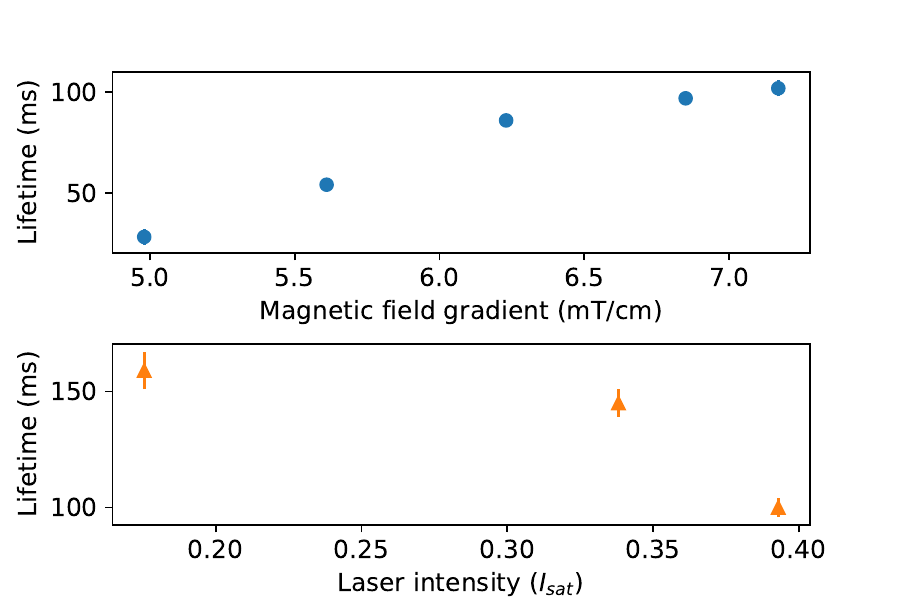}%
	\caption{Lifetime in the $^1S_0 \rightarrow {^1P_1}$ MOT as a function of magnetic field gradient with $s=0.4$ and  laser intensity with $dB/dz = 7\,\mathrm{mT/cm}$. The vacuum pressure is $5\cdot10^{-9}\,\mathrm{mbar}$ corresponding to a background-gas-collision limited lifetime of $200\,\mathrm{ms}$.  \label{fig: Lifetime}}
\end{figure}
\begin{figure}
	\includegraphics[width=0.9\linewidth]{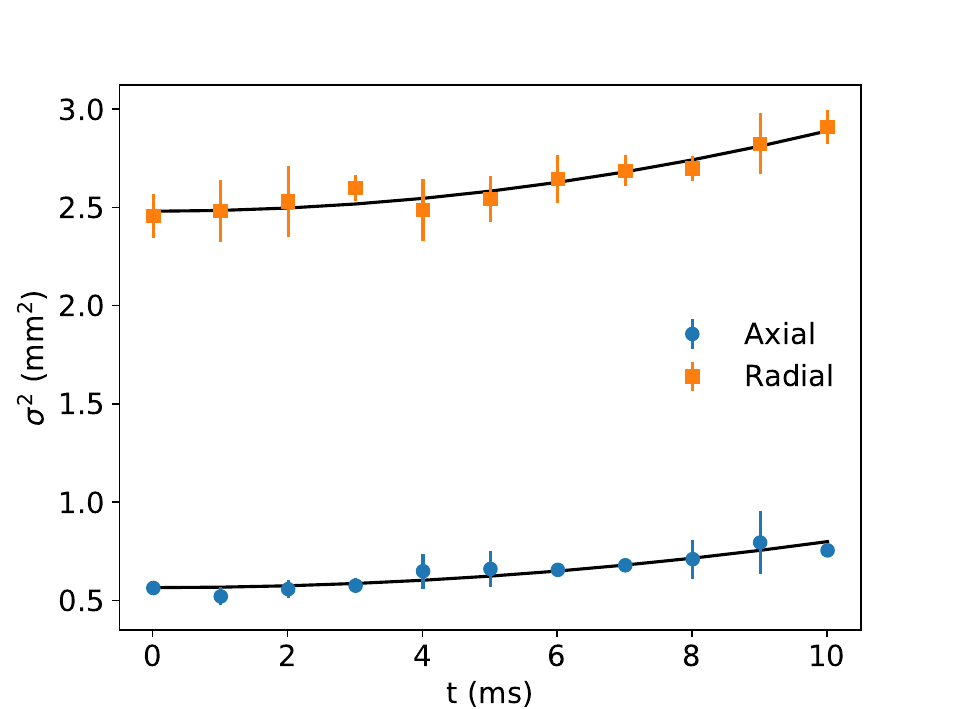}%
	\caption{Expansion of the atomic cloud in axial (blue) and radial (orange) direction with time after the broadband MOT on the $^1$S$_0 \rightarrow ^3$P$_1$ transition. The fit yields $T_{ax}=(23\pm 3)\,\mu\mathrm{K}$ and $T_{rad}= (43\pm 4)\,\mu\mathrm{K}$. \label{fig:Temperature_red}}
\end{figure}
\begin{figure}
	\includegraphics[width=0.95\linewidth]{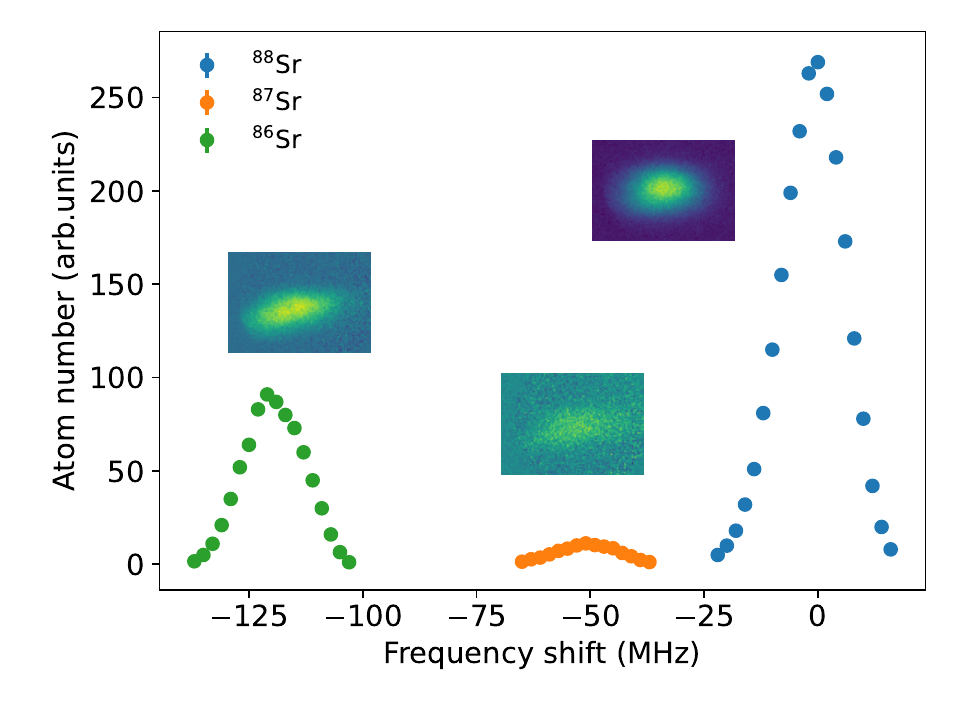}%
	\caption{Atom number as a function of frequency detuning for $^{88}$Sr, $^{87}$Sr and $^{86}$Sr. Insets show fluorescence images for the atom clouds. For $^{86}$Sr and $^{87}$Sr, multiple images are averaged and the averaged background substracted to obtain an improved signal to noise. Further, the exposure time is increased from $100\,\mu\mathrm{s}$ to $300\,\mu\mathrm{s}$. Figure adapted with permission from S. Bondza \cite{bon23a}. \label{fig:Spectrum}}
\end{figure}

We further analyzed the first cooling stage with respect to life time. The lifetime $\tau$ can be determined from a loading curve with $N_{atoms}(t)=r\tau(1-e^{-t/\tau})$ where $r$ is the loading rate and $N_{atoms}$ the atom number \cite{sit21}.  We found that the life time shows a dependency on laser intensity as well as magnetic field gradient as presented in Fig. \ref{fig: Lifetime}. \\
We observe the lifetime of the atoms in the MOT to decrease with increasing intensity. One possible explanation lies in the strong increase in temperature. Fast atoms are more likely to be able to escape the trap while in a dark state before being able to be repumped resulting in a loss channel. We observed that decreasing the power in the repumping beams decreases the atom number which supports this explanation.
The lifetime increases with increasing magnetic field gradient as previously observed in the grating MOT as well \cite{sit21, bon22}.
We measured a maximal lifetime of $\tau_{tot}=160\,\mathrm{ms}$ at a vacuum pressure of $5\cdot10^{-9}\,\mathrm{mbar}$ at a laser intensity $s=0.18$. At $5\cdot10^{-9}\,\mathrm{mbar}$, background gas collisions limit the MOT lifetime to $\approx200\,\mathrm{ms}$ \cite{nag03}. 
In our set-up outgassing from the strontium oven limits the vacuum pressure.
Using $\frac{1}{\tau_{vacuum} } + \frac{1}{\tau_{FMOT}} = \frac{1}{\tau_{total}}$, we calculate a lifetime of the Fresnel MOT of up to $\tau_{FMOT} = 800\,\mathrm{ms}$.\\
We transfer the pre-cooled $^{88}$Sr atoms to a broadband MOT operated on the $^1$S$_0 \rightarrow ^3$P$_1$ transition at $689\,\mathrm{nm}$ where the laser was artificially broadened to about $1\,\mathrm{MHz}$. This procedure is common and improves transfer efficiency as a temperature in the milli-Kelvin range corresponds to a Doppler shift on the order of megahertz\cite{kat99}. We achieve a transfer efficiency of $50\,\%$, comparable to typical transfer efficiencies in the six-beam geometry. Here, the atoms where further cooled to a temperature of $(23\pm 3)\,\mu\mathrm{K}$ in axial and $(43\pm 4)\,\mu\mathrm{K}$ in radial direction, respectively, as seen in Fig.~\ref{fig:Temperature_red}. This result presents a first demonstration of multi-color cooling with the novel structure.
As a next step, atoms can be loaded into a dipole trap for further cooling. Trap depths for optical lattices typically reach few hundred recoil energies corresponding to tens of microkelvins \cite{hob20a}.
To optimize transfer efficiency to a dipole trap, atoms are commonly further pre-cooled on the $^1$S$_0 \rightarrow ^3$P$_1$ transition in a so-called single-frequency MOT where narrow-line cooling laser is used. An unmodulated, narrow-line, single-frequency grating MOT with a similar geometry has already been demonstrated for strontium, reaching temperatures below $10~\mu K$ \cite{bon22, elg22a}.\\
Apart from $^{88}$Sr, we also trapped and cooled $^{86}$Sr as well as the fermionic isotope $^{87}$Sr on the $^1$S$_0 \rightarrow ^1$P$_1$ transition. We observe an axial shift between the bosonic and fermionic MOT which is consistent with a theory of the trapping dynamics \cite{bar23} and previous observations by other groups \cite{elg22a}.\\
To improve the atom number for the $^{86}$Sr and $^{87}$Sr MOTs, we increase the temperature of the thermal atom source to $400\,$°C, thus generating a higher atom flux. Fig.~\ref{fig:Spectrum} shows the relative atom number as a function of detuning of the cooling laser for $^{86}$Sr, $^{87}$Sr, and $^{88}$Sr where $0\,\mathrm{MHz}$ detuning is chosen as the frequency yielding the maximum atom number for $^{88}$Sr. The frequency shifts for $^{86}$Sr and $^{87}$Sr correspond to the isotope shifts of the $^1S_0 \rightarrow {^1P_1}$ transition \cite{lor83}. \\
While $^{86}$Sr and $^{87}$Sr have a similar abundance, we trap about four times as much $^{86}$Sr as $^{87}$Sr. One possible reason lies in the loss channel due to insufficient repumping for fast atoms impacting fermionic strontium more severely, as the repumping lasers need to be frequency modulated to cover all Zeeman sublevels which reduces the saturation parameter. To increase the atom number for $^{87}$Sr and the signal to noise ratio, more efficient loading geometries can be employed, e.\,g., including a Zeeman slower. 

\section{Conclusion}
We have demonstrated two-stage laser cooling of bosonic $^{88}$Sr to micro-Kelvin temperatures and trapping of $^{86}$Sr and fermionic $^{87}$Sr in the first MOT stage with a novel, quasi-planar, achromatic reflector in a single-beam geometry. The Fresnel reflector is a low-cost, easy to manufacture device without the need for microfabrication technologies. It further combines the advantages of the grating MOT offering unrestricted radial access and a high level of miniaturization with the achromatic beam path of the tetrahedral MOT. This enables trapping within a fixed volume in an ideal tetrahedral geometry independent of the wavelengths, thus allowing for optimized trap depths which is critical in particular for fermionic alkaline-earth-like isotopes.
As a result, it promises new possibilities for robust, multi-color cooling of one or multiple atomic species.\\
We expect that the performance of the Fresnel MOT can be further improved by addressing sources of spatial intensity fluctuations, such as uneven reflectivity of the coating and imperfect edges of neighbouring surfaces, employing differential pumping to reduce the vacuum pressure, and a more sophisticated loading scheme increasing atom number and reducing background fluorescence. 
Our work is an important step in the widespread adoption of single-beam MOT designs based on planar structures for quantum sensors aiding the essential reduction of their SWaP.

\section{Acknowledgement}
We thank Stephan Metschke and his group for the manufacturing of the Fresnel structure and Andre Felgner for the characterization of its surface. We further thank Sören Dörscher for careful reading of the manuscript and helpful discussion.
This work is financially supported by the State
of Lower-Saxony through the VW Vorab. We further acknowledge support by the Deutsche Forschungsgemeinschaft (DFG, German Research Foundation) under Germany’s Excellence Strategy—EXC-2123 QuantumFrontiers—Project-
ID 390837967.

\section*{Author Declarations}

\subsection*{Conflict of Interest}
SB, TL and CL are inventors of Patent DE: 10 2020 102 222.0. PTB holds the rights to this patent. 

\subsection*{Author Contributions}

\textbf{Saskia Bondza}: Conceptualization, Investigation (lead), Methodology, Data Analysis, Visualization, Writing/Original Draft Preparation, Review and Editing (lead)
\textbf{Tobias Leopold}: Conceptualization, Methodology, Supervision
\textbf{Roman Schwarz}: Investigation, Writing/Original Draft Preparation, Review and Editing
\textbf{Christian Lisdat}: Conceptualization (lead), Supervision, Visualization,  Writing/Original Draft Preparation, Review and Editing

\section*{Data Availability}

The data that support the findings of this study are available from the corresponding author upon reasonable request.

%

\end{document}